# Arbitrary fractional quantization in Dirac systems

Christos Papapanos[1], Rushin Contractor[1], Mariusz Drong[1], Matteo Seclì[1], Boubacar Kanté[1,2,†]

[1]*Department of Electrical Engineering and Computer Science, University of California Berkeley, Berkeley, CA, 94720, USA*
[2]*Material Science Division, Lawrence Berkeley National Laboratory, Berkeley, CA, USA*

[†]bkante@berkeley.edu

**Oscillations are ubiquitous wave phenomena in physical systems ranging from electromagnetic and acoustic to gravitational waves. The behavior of finite-size systems is traditionally understood to be governed by fundamental oscillatory modes arising from bulk physics and boundary conditions. A paradigmatic example is the particle-in-a-box model introduced with the advent of quantum mechanics, in which confinement leads to discrete resonances and quantized energy levels. Such quantization underpins phenomena including semiconductor quantum dots, where electronic waves are confined in all three spatial dimensions, producing standing-wave modes analogous to the vibrational states of a guitar string. These modes are characterized by integer quantum numbers corresponding to the number of envelope oscillations fitting within the cavity. Recently, counter-intuitive modes have been observed in finite systems with Dirac dispersion, including states that do not oscillate spatially, yet a general theoretical framework for modes in such cavities has been lacking. Here, we discover the phenomenon of arbitrary fractional quantization in wave physics and show that, in finite-size boxes with linear dispersion, the quantum number need not be an integer but can take any real value, including zero. Using Bloch theory in a Dirac-cone photonic crystal with a controllable fractional number of unit cells at its boundaries, we demonstrate continuous control of the cavity-mode envelope wavenumber. We introduce a unified nomenclature for these unconventional modes and derive their corresponding wavefunctions. These counter-intuitive states in open Dirac potentials challenge conventional notions of quantization and open new avenues across wave physics.**

Cavity modes are fundamental in wave physics because they represent the countable patterns of standing waves that can exist within a confined structure often referred to as a cavity, box, or potential well. These modes arise due to the combined effects of the bulk-physics of the system and boundary conditions, allowing only specific wavelengths, and thus frequencies, to resonate within the cavity [2]. Quantization is central to a wide range of physical phenomena and technologies including in lasers where only certain optical modes can build up coherently [3,4], in quantum mechanics where atoms exhibit discrete energy levels [5-7], and in cavity quantum electrodynamics where the cavity determines how photons interact with matter [8-10]. In crystalline structures of finite size, the curvature of bands plays the role of mass, and the physics of the formation of discrete modes or quantization is essentially the same as in ordinary physical systems such as guitar strings. This is illustrated in Fig. 1a based on a crystal with a conventional quadratic dispersion where two bands are separated in frequency at the band edge by a finite size band gap. When truncated to form a potential well of length $D$, be it electronic, acoustic, or photonic, the modes originating from the quantization of the bands are usually conventional oscillating modes in a potential well at finite wave vectors $k = n\pi/D$ where $n = 1, 2, 3, ...$ [11]. The integer $n$ quantifies the number of half-envelope oscillations that can fit in the box.

In certain crystalline systems, such as graphene, the energy-momentum relation becomes linear, leading to properties of massless relativistic particles [12,13]. In photonic crystals [14,15], such linear dispersion can be systematically implemented using symmetry considerations, resulting in the formation of Dirac-cones in the band structure [16-21]. In Fig. 1, a Dirac-cone is obtained when the frequency separation between the two bands of Fig. 1a (top) approaches zero, forming a linear dispersion governed by equation Eq. 1. The bands that were purely odd or even in the presence of a band gap become mixed. Cavity modes of the corresponding finite size systems (Fig. 1b and Fig. 1c) are thus a linear combination of the odd ($|-\rangle$) [yellow component] and even ($|+\rangle$) [blue component] bands. Potential wells using such a linear dispersion were named open-Dirac potentials [1]. Modes in an open-Dirac potential are denoted $P_{|n|,\Pi}$ where $P$ is the position of the band compared to the frequency of the Dirac-point, i.e., lower ($L$) or upper ($U$), $|n|$ is the absolute value of the quantization number, and $\Pi$ is the total mode parity [even (+) or odd (-)]. Figures 1b,c show that the quantization in one dimensional open-Dirac potentials is surprisingly different from the conventionally understood physical systems. Not only is a mode at zero wavenumber permitted, but the quantum number in general does not need to be an integer and can instead be any arbitrary real number, enabling arbitrary fractional quantization. We will show that the quantization is governed by two quantities including the admixture of the odd and even bands when the linear dispersion is formed in the bulk of the crystal and the fraction of unit-cells at the boundary of the finite sized cavity. Unlike atomic systems in which an atom at the edge cannot in general be truncated arbitrarily, the unit-cells of photonic crystals can be truncated at arbitrary positions to leave a fraction of unit-cell at the edge (FUCE). The linear dispersion and the FUCE at the boundary of the potential enable unconventional modes, never observed before in any wave-physics system to the best of our knowledge. Figure 1b (top) illustrates an open-Dirac potential with what we denote "anomalous integer quantization" where the quantum number of cavity modes can be zero with an unconventional mode at $k \sim 0$ for certain boundary termination. Modes thus exist at $k = n\pi/D$, where n is an integer and the formation of mode follows an anomalous integer quantization. Figure 1b (bottom) presents the wavefunctions for modes $L_{0,+}$, $L_{1,-}$, $L_{2,+}$, $U_{1,-}$, and $U_{2,+}$. Mode $L_{0,+}$ at $k \sim 0$ does not oscillate in the open-Dirac potential. Figure 1c (top) presents an open-Dirac potential with what we denote "arbitrary fractional quantization" where the quantum number of cavity modes can now take arbitrary real values. Modes exist at $k = (m + \varepsilon)\pi/D, \varepsilon \in [0,1)$ where the quantum number can now take arbitrary real values, and the formation of modes follows an arbitrary continuously tunable fractional quantization. Figure 1c (bottom) illustrates wavefunctions with $\varepsilon = 1/e$ giving rise to modes $L_{1-e^{-1},+}$, $L_{2-e^{-1},-}$, $U_{e^{-1},-}$ and $U_{1-e^{-1},+}$.

To understand the formation of these unconventional modes, solutions of the wave equation are searched in the form $h_{k,m}(y) e^{iky}$, where $m$ denotes the Bloch modes symmetry [16,17]. In the appropriate basis near the high symmetry Γ-point, we can consider that only two degenerate or nearly degenerate modes exist [22]. Since these modes form a complete basis around the Γ-point, a parity can be assigned with an odd function $h_{k,-1}(y)$ and even function to $h_{k,+1}(y)$. Applying first order perturbation theory [16,22], an effective momentum-space Hamiltonian operator around the Γ-point in the unperturbed parity basis $|\pm\rangle$ is defined with $\langle y|\pm\rangle = h_{0,\pm}(y) = h_\pm(y)$. The resulting Hamiltonian operator provides a general framework applicable to other wave-physics systems while allowing direct analysis in the frequency/energy domain,

$$\widehat{H}_k = \widehat{1}\omega_0 + \hat{\sigma}_z \delta + \hat{\sigma}_y v_g k = \begin{pmatrix} \omega_0 + \delta & -i v_g k \\ i v_g k & \omega_0 - \delta \end{pmatrix} \quad (1)$$

where $\widehat{1}$ is a unity matrix, $\hat{\sigma}_y$ and $\hat{\sigma}_z$ are Pauli matrices and the group velocity $v_g$ around the Γ-point is taken to be real without any loss of generality, i.e., up to a unitary transformation and, δ is a small frequency walk-off from the degeneracy [22]. Diagonalizing $\widehat{H}_k$ then yields both the energy dispersion of the upper/lower bands ($b = \{u, l\}$, respectively)

$$\omega_b^{\Delta}(k) = \omega_0 \pm |v_g k| \sqrt{1 + \left(\frac{\Delta}{k}\right)^2} \quad (2)$$

and the eigenvectors

$$|b_k^{\Delta}\rangle \propto |+\rangle - ic_b(\Delta, k)|-\rangle, \quad (3)$$

where $\Delta = \delta/v_g$, $c_u(\Delta, k) = \Delta/k + sgn(k)\sqrt{1 + (\Delta/k)^2}$ and $c_l(\Delta, k) = \Delta/k - sgn(k)\sqrt{1 + (\Delta/k)^2}$ have the property $c_b(\Delta, -k) = -c_b(\Delta, k)$. For $k = 0$, by definition $|u_0^{\Delta}\rangle \sim |-\rangle$ and $|l_0^{\Delta}\rangle \sim |+\rangle$, i.e. the eigenmodes are always purely even or purely odd at the Γ-point. As $\Delta \to 0$ the eigenvalues of $\widehat{H}_k$ in Eq. (2) become linearly dependent on k and result in a Dirac cone, while the modes $|b_k^{\Delta}\rangle$ become a bona fide admixture of both even and odd modes $|+\rangle \pm i|-\rangle$, as shown in Figs.1b,c (top) by the green lines. From now on, we drop the index Δ as only linear dispersion where the gap is closed (Δ=0), will be considered. The eigenfunctions of the effective Hamiltonian $\widehat{H}_k$ satisfy $|b_{-k}\rangle = |b_k\rangle^*$ [17].

The designed two-dimensional (2D) photonic crystal (PhC) features a triangular lattice. It consists of a material with a refractive index of 3.4, similar to that of InGaAsP [1], and air holes of radii $r \simeq 0.211a$ (for the quadratic dispersion) and $r \simeq 0.222a$ (for the linear dispersion) where *a* is the periodicity. The dispersion around the Γ-point is well described on a basis of only three PhC modes, labeled $B_1$, $E_{2a}$ and $E_{2b}$ according to their symmetry classification [18]. The dispersion relations are calculated using the finite element method, and results are presented in Fig. 2a for the quadratic dispersion. At the Γ-point, the $B_1$ mode and the two $E_2$ modes are non-degenerate, and they remain unmixed for wavevectors very close to the Γ-point. Consequently, distinct colors are assigned to each dispersion line based on the $H_z$ field distribution at the Γ point (see inset). Notably, the $E_{2a}$ and the $E_{2b}$ modes switch position when the momentum direction changes from ΓM to ΓK, and vice versa. When the unit-cell is perturbed to achieve linear dispersion (Fig. 2b), the modes forming the Dirac cone become mixed. Specifically, the $B_1$ mode (blue) and the $E_{2a}/E_{2b}$ modes (yellow/red) mix to form the $B_1 \pm iE_{2a}$ (blue+yellow=green) dispersion lines along the ΓM direction and the $B_1 \pm iE_{2b}$ (blue+red=purple) dispersion lines along the ΓK direction. It is worth noting that the $B_1$ and the $E_{2a}$ modes have the same/opposite symmetry along the ΓM/ΓK respectively, giving rise to the designations "even" ($|+\rangle$) for $B_1$ and "odd" ($|-\rangle$) for $E_{2a}$ modes.

Full translational invariance in the PhC plane, i.e. periodic boundary conditions (PBCs) in both directions, was so far assumed. To investigate finite structures, the PhC is truncated in one direction to form a supercell of length $D$, also called box, cavity, or potential. In the potential, multiple unit-cells are connected along, e.g. the ΓM direction, while periodic boundary conditions (PBCs) are applied in the ΓK direction. Figure 2c illustrates a potential comprising a few unit-cells (six unit-cells for visualization). However, a potential is not necessarily constrained to any number of unit-cells and could in principle have a non-integer number of unit-cells along its length. For instance, in Fig. 2c, an additional quarter of unit-cell has been added to both ends of the potential. More generally, any fractional unit-cell can be added at the edges of potentials, a new parameter that we call *Fraction of Unit-Cell at the Edge* (FUCE). Since the cavity (potential) is finite in the ΓM or $\hat{y}$-direction, and its interfaces at $|y| = D/2$ cause reflections of waves inside the cavity, we look for bound states in the form of counter-propagating Bloch waves. The constraint that the energy needs to decay in time leads the upper band to have $k > 0$ and the lower band to have $k < 0$ (see supplementary section 1):

$$|U_{k,\pm}\rangle = |b_k\rangle e^{i\tilde{k}y} \pm |b_{-k}\rangle e^{-i\tilde{k}y}, (upper\ band) \quad (4a)$$

$$|L_{k,\pm}\rangle = |b_{-k}\rangle e^{i\tilde{k}y} \pm |b_k\rangle e^{-i\tilde{k}y}, (lower\ band) \quad (4b)$$

where, $U$ and $L$ denotes the band at higher $U$ and lower $L$ energy, the $\pm$ denotes the overall parity of the solution, and $\tilde{k} = k - i\kappa$ is the complex wavenumber [23]. Because the $B_1$ and the $E_{2a}$ modes are symmetry-mismatched with plane waves, losses can be neglected at edges. The upper band modes are therefore:

$$|U_{k,+}\rangle \propto \cos(k\ y)|+\rangle + \sin(k\ y)|-\rangle, \quad (5a)$$

$$|U_{k,-}\rangle \propto \sin(k\ y)|+\rangle - \cos(k\ y)|-\rangle \quad (5b)$$

and the lower band ones, including the sign of k, are:

$$|L_{k,+}\rangle \propto \cos(k\ y)|+\rangle - \sin(k\ y)|-\rangle, \quad (6a)$$

$$|L_{k,-}\rangle \propto \sin(k\ y)|+\rangle + \cos(k\ y)|-\rangle \quad (6b)$$

Since these modes are resonant, we need to ensure, by definition, they are invariant upon a complete cavity round-trip, consisting in two reflections at the cavity edges and two propagations throughout the full cavity length. Each reflection adds a phase $2\theta(\eta) = 2Arg\{b_k(\eta)\}$, different for each FUCE value ($\eta$), thus, the resonance condition for the upper band is:

$$kD = m\pi - 2\theta(\eta), \quad (7)$$

where $m$ is an integer number. Similar reasoning gives the same results for the lower band. The mechanism that allows for continuous tuning of envelope wavenumbers in open-Dirac cavities can be explained by the Bloch theorem and the phase resonance condition. When Bloch waves propagate along the cavity, they acquire an envelope (round-trip) phase that depends on the overall length of the cavity. However, they also acquire a unit-cell-level phase, which resets at every integer unit-cell traveled due to periodicity. The overall phase must be an integer multiple of $2\pi$ to form a resonant mode. For conventional cavities (with quadratic dispersion and no band mixing), the overall phase coincides with the round-trip phase, and the envelope wavenumber must therefore be an integer multiple of $\pi/D$ regardless of the FUCE. However, the total number of unit-cells in our PhC cavity with linear dispersion does not necessarily need to be an integer. In this case, the wave propagation in fractional unit-cells at the boundaries results in a potentially non-zero unit-cell phase $2\theta(\eta) = 2Arg\{|b_k\rangle\}$. This additional phase must be compensated by a shift in envelope wavenumber for the resonant condition to hold, leading to a generalized resonance condition $k = m\pi/D - 2\theta/D$. Notably, the generalized resonance condition allows for an envelope wavenumber that is arbitrarily controllable. It is worth noting that the generalized resonance condition gives results equivalent to the Maxwell equations approach (see section 4 of the Supplementary information).

The continuous control of envelope wavenumber is enabled by the specific band mixing $B_1 \pm iE_{2a}$ in the Dirac-cone dispersion, leading to delocalized (massless) nature of unit-cell modes as opposed to localized (massive) pure unit-cell modes in the quadratic dispersion. The mixed Dirac-cone unit-cell modes behave as travelling waves and thus the unit-cell can be terminated anywhere (see section 6 of the Supplementary information). The FUCE value used to construct the open-Dirac potential (cavity) determines the envelope's wavenumber (Fig. 2d), whose absolute value is any number between zero and one. In Fig. 2d, the wavenumbers extracted from finite element simulation are overlaid with theoretical predictions using Eq. 7, where the even and odd symmetry modes across the full range of FUCE values are also shown with blue squares and green triangles respectively. Due to the unit-cell symmetry, FUCE values greater than 0.5 are equivalent to those between 0 and 0.5.

To further illustrate the impact of the FUCE on the quantization of the bands, we discuss two cases corresponding to "anomalous integer quantization" (Fig. 3) with $FUCE = 0.24$, and to "arbitrary fractional quantization" (Fig. 4) with $FUCE = 0.13$. In Fig. 3a, the dispersion of the PhC (infinite structure) along the ΓM direction is overlaid with the quantized modes of a finite structure. An excellent agreement is observed with envelope wavenumbers taking integer values, including zero.

Eqs. (5) and (6) indicate that each cavity mode can be decomposed into a linear combination of the $B_1$ and the $E_{2a}$ basis modes. Figures 3b-i present the projection of selected cavity modes into their $B_1$ (continuous lines) and $E_{2a}$ (dashed lines) components across the cavity for eight of the lowest order cavity modes shown in Fig. 3a as black markers. The total parity of the modes is the product of the unit-cells' and the envelope's symmetries and is represented by blue and green for even and odd modes respectively. A cavity mode exhibiting even/odd symmetry, has maximum $B_1$ /$E_{2a}$ components at the center of the cavity. Modes in Figs. 3b-e belong to the lower (L) band, while modes in Figs. 3f-i are from the upper (U) band. The mode denomination, using the proposed nomenclature, is provided in each plot. For example, Fig. 3b presents the $L_{5,-}$ mode, characterized by a wavenumber $k = 5\frac{\pi}{D}$ belonging to the lower (L) band. A particular feature of the anomalous integer quantization is evident from Fig. 3e, where a flat mode in real space, with $k = 0\frac{\pi}{D}$ is predicted theoretically and confirmed through simulations. In this case, only the $B_1$ component is present, with no mode mixing at the Γ-point, as previously discussed. Figure 4a presents cavity modes overlaid with the linear dispersion for a FUCE giving rise to an arbitrary fractional quantization (FUCE=0.13). Figures 4b-i present the projection of selected cavity modes into their $B_1$ (continuous lines) and $E_{2a}$ (dashed lines) components. Interestingly, the $B_1$ and the $E_{2a}$ components do not vanish at the cavity edges unlike in the anomalous integer quantization. Figures 3 and 4 also indicate that while the band are equally mixed in the infinite crystal, quantization leads to a position dependent mode mixing in open-Dirac potentials with energy oscillating between the two components of the cavity modes.

To further confirm that a position dependent mode mixing occurs between unit-cell modes ($B_1$ and $E_{2a}$) in open Dirac-potentials (see inset in Fig. 2a), Fig. 5 present three selected modes, their decomposition in the unit-cell basis and the near-field distribution inside the cavities. Figure 5a (top) indicates that the $L_{0,+}$ mode only has $B_1$ component and no $E_{2a}$ components. A close look at Fig. 5a confirms that the near field of the cavity only has $B_1$ field pattern. In Fig. 5b (top) presenting mode $L_{3,-}$, the left edge of the cavity (cyan region) only contains $B_1$ component while the bulk (orange region) contains both $B_1$ and $E_{2a}$ components. This is confirmed by a close look at the near field pattern (bottom). Finally, Fig. 5c (top) indicates that the $U_{2.67,-}$ mode contains both $B_1$ and $E_{2a}$ components at the edge of the cavity (cyan region) and only the $E_{2a}$ component in the bulk highlighted orange region. This is also confirmed by the near field distribution in the cavity.

We have thus discovered a new form of quantization and unconventional wave functions arising in wave-physics systems with Dirac dispersion. The formation of cavity modes follows an arbitrary fractional quantization rather than the conventional integer quantization. We introduced a novel degree of freedom called fraction of unit-cell at the edge (FUCE) of the potential. The new degree of freedom terminates a potential with a fraction of an "atom" or "artificial atom" in engineered wave-physics systems such as photonic crystals. In systems with linear dispersion, the tunable FUCE enables unprecedented control over mode envelopes through resonances that intertwine macroscopic envelopes and microscopic unit-cells degrees of freedom. These surprising results could reshape our understanding across classical and quantum wave-physics systems.

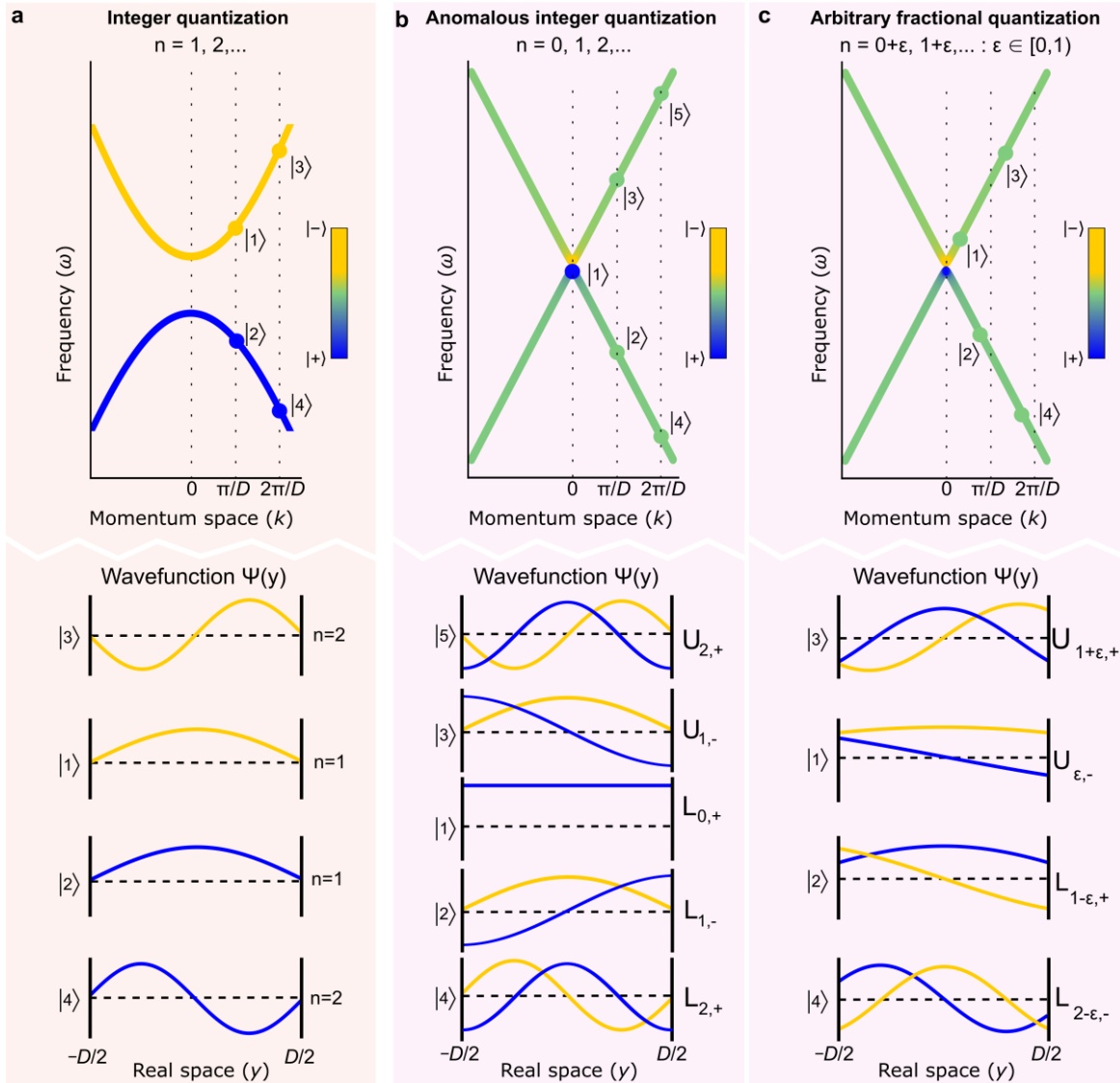

**Fig. 1 | Conventional rectangular potential and open-Dirac potential. a**, (Top) Conventional quadratic band structure when an odd (yellow) and an even (blue) bands are separated in frequency at the band edge by a finite size band gap. (Bottom) When confined within a potential well, be it electronic, acoustic, or photonic, the modes, originating from the quantization of the bands are conventional oscillating modes in a potential well at finite wavenumbers $k = n\pi/D$ where $n = 1,2,3,…$ [11]. (**b, c**) As the frequency separation between the two bands approaches zero to form a Dirac cone, the dispersion governed by Eq. 2 turns linear and the bands are now mixed and no longer purely odd or even. The cavity modes of an open-Dirac potential with linear dispersion are thus a linear combination of the odd ($|-\rangle$) [yellow component] and even ($|+\rangle$) [blue component] bands. A potential well using such a linear dispersion is an open-Dirac

potential. Modes in an open-Dirac potential are denoted $P_{|n|,\Pi}$ where $P$ is the position of the band compared to the frequency of the Dirac-point, i.e., lower (L) or upper (U), $|n|$ is the absolute value of the quantization number, and $\Pi$ is the total mode parity [even (+) or odd (-)]. **b**, (Top) Open-Dirac potential with anomalous integer quantization where the quantum number of cavity modes can be zero with an unconventional mode at $k\sim 0$ for certain boundary conditions. (Bottom) Wavefunctions for mode $L_{0,+}$, mode $L_{1,-}$ mode $L_{2,+}$, mode $U_{1,-}$, and $U_{2,+}$. **c**, (Top) Open-Dirac potential with arbitrary fractional quantization where the quantum number of cavity modes can now take arbitrary real values. Modes exist at $k = (m+\varepsilon)\pi/D, \varepsilon \in [0,1)$, where the quantum number $n = m + \varepsilon$ (where $m$ is an integer) can now take arbitrary real values, and the formation of modes follows an arbitrary continuously tunable fractional quantization. (Bottom) Example of wavefunctions with $\varepsilon = 1/e$ giving rise to modes $L_{1-e^{-1},+}$, mode $L_{2-e^{-1},-}$, mode $U_{e^{-1},-}$ and mode $U_{1+e^{-1},+}$.

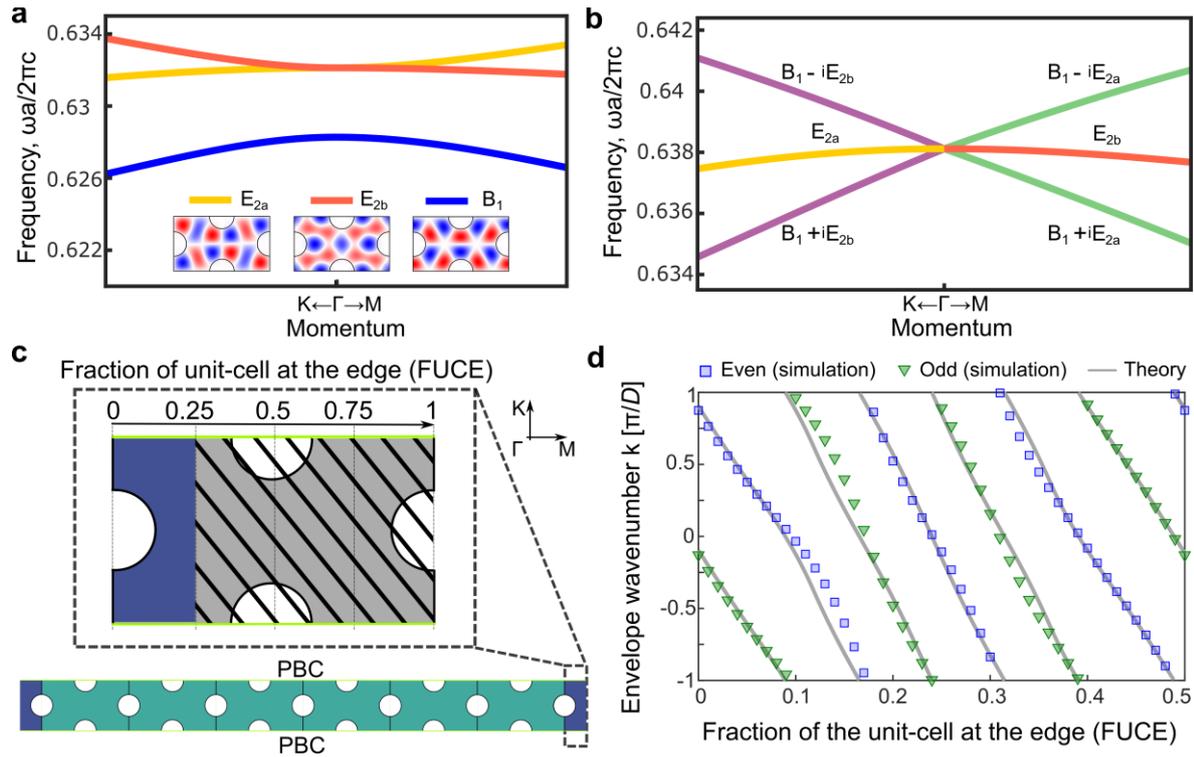

**Fig. 2 | Photonic-crystal implementation of an open-Dirac potential, fractional boundaries, and modes formation mechanism. a**, Quadratic dispersion for bands in a $C_{6v}$ symmetry photonic crystal when the band originating from the $B_1$ and $E_2$ symmetry groups are non-degenerate at the center of the Brillouin zone (Γ-point). Inset shows the out-of-plane magnetic field distribution in the unit-cell at the Γ-point. **b**, The dispersion turns linear when the $B_1$ and $E_2$ bands are tuned to a degeneracy. Note that the $E_{2a}$ couples with the $B_1$ mode in the ΓM direction (green lines) whereas the $E_{2b}$ mode couples with the $B_1$ mode in the ΓK direction (purple lines). **c**, The unit-cell of the photonic crystal under consideration which can be used to construct 1D potential wells (named supercells or cavities) that are of finite size in the ΓM direction and infinite in the ΓK direction with periodic boundary conditions (PBC). A supercell, of total length D, can be terminated by a non-integer number of unit-cells (fraction of unit-cell) at the edge. The fraction of unit-cell at the edge (FUCE) of the potential well alters the quantization in the potential well (see Fig. 2d). The illustrated 1D supercell is composed of six unit-cells, supplemented with a fractional unit-cell at the edge (dark purple for visualization) of 0.25. The dashed gray area is not used in the supercell termination and is shown for better understanding of the different FUCE values. The edges of the supercell are perpendicular to the ΓM direction giving rise to cavity modes originating from the bands $B_1$ and $E_{2a}$ (green lines in Fig. 2b), $E_{2b}$ (red line in Fig. 2b) of the infinite crystal. **d**, Wavenumber of the envelope for lowest order even (blue) and odd (green) modes as a function of the FUCE. The markers represent envelopes extracted from finite element simulations and the grey continuous lines are the

prediction of our theory (see Eq. 7). Interestingly, the wavenumber of the envelope of cavity modes can be continuously tuned.

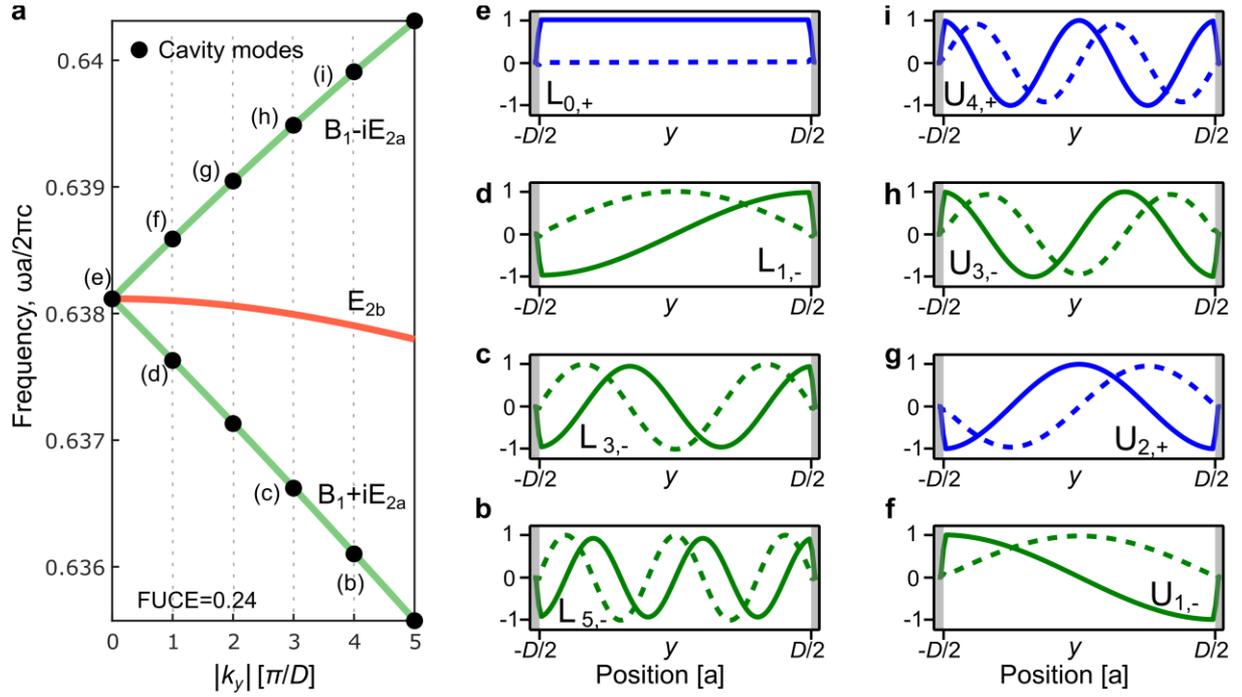

**Fig. 3 | Anomalous integer quantization in an open-Dirac potential**. **a**, The frequency of the cavity modes with cavity size $N = 100a\sqrt{3}$, where $a$ is the size of the unit-cell, overlaid on the dispersion of the infinite photonic crystal along the ΓM (or y) direction for $FUCE = 0.24$. This FUCE leads to quantum numbers for cavity modes of $|n| = 0,1,2,3,4$, etc. **b-i**, Projection of the cavity modes into their $B_1$ (continuous lines) and $E_{2a}$ (dashed lines) components across the cavity for eight of the lowest order cavity modes shown in Fig. 3a as black markers. The parity of the modes is the product of the unit-cells and the envelope's symmetries. In Figs (b-i), the result of this product is shown by the color, green/blue for odd/even parity. When the $B_1(E_{2a})$ mode is maximum in the center, the total parity of the mode is even (odd). Note that there is no $E_{2b}$ component as this band does not couple to the $B_1$ band in the direction of finiteness of the cavity (quantization direction) that is the direction ΓM as shown in Fig. 3a. All observed modes are unusual, and we also observe a flat-envelope mode with $n = 0$ in Fig. 3e where only the $B_1$ mode is present throughout the cavity aperture.

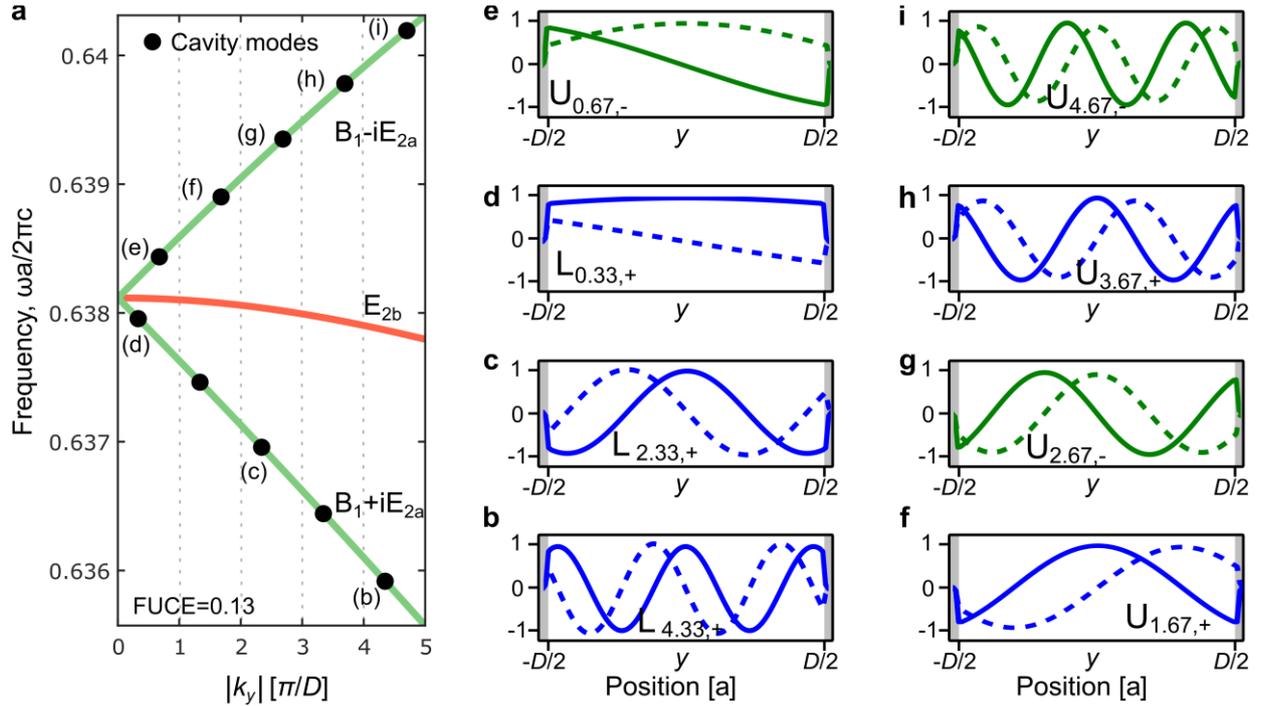

**Fig. 4 | Arbitrary fractional quantization in an open-Dirac potential. a**, The frequency of the cavity modes with cavity size $N = 100a\sqrt{3}$, where $a$ is the size of the unit-cell, overlaid on the dispersion of the infinite photonic crystal along the ΓM (or y) direction for $FUCE = 0.13$. This FUCE leads to quantum numbers for cavity modes of $|n| = 0.33, 0.67, 1.33, 1.67$, etc. **b-i**, Projection of the cavity modes into their $B_1$ (continuous lines) and $E_{2a}$ (dashed lines) components across the cavity for eight of the lowest order cavity modes shown in Fig. 3a. The parity of the modes is the product of the unit-cells and the envelope's symmetries. In Figs. (b-i), the result of this product is shown by the color, green/blue for odd/even parity. Note that there is no $E_{2b}$ component as this band does not couple to the $B_1$ band in the direction of finiteness of the cavity (quantization direction) that is the direction ΓM as shown in Fig. 3a.

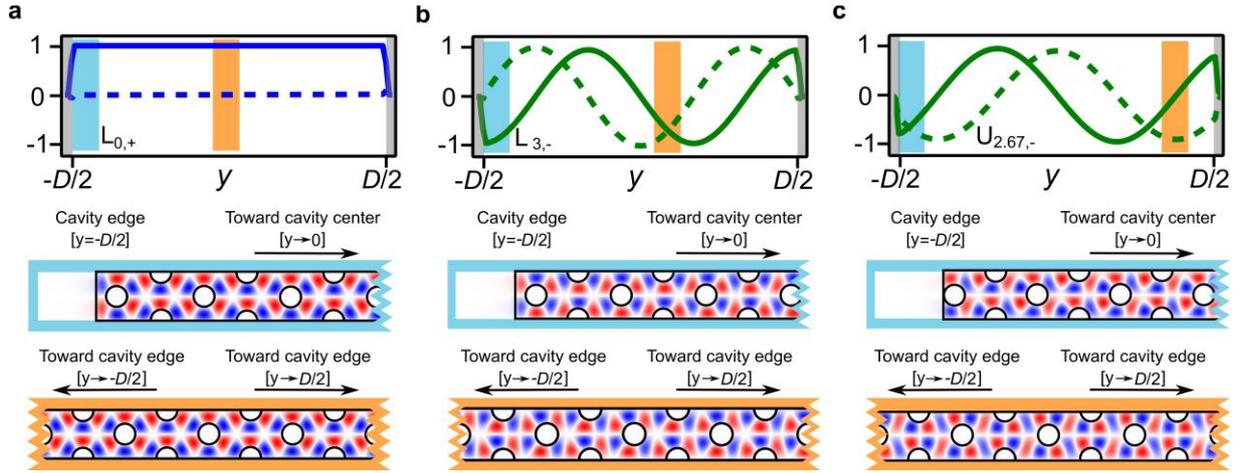

**Fig. 5 | Near-field evidence of position dependent mode mixing in open-Dirac potentials.** A close look at cavity modes shows the mixing of the $B_1$ and the $E_{2a}$ unit-cell modes (see inset in Fig. 2a) at different positions of the cavities indicated by the cyan or orange colors. The continuous line is the $B_1$ component and the dashed line is the $E_{2a}$ component. Three cases are shown with $FUCE = 0.24$ for Fig. 5a-b and $FUCE = 0.13$ for Fig. 5c. **a**, $FUCE = 0.24$ and the $L_{0,+}$ cavity mode (same as in Fig. 3e). This is the unconventional flat mode where $n = 0$ and only the $B_1$ mode exists in the entire cavity as confirmed by the near-field coinciding with $B_1$ mode around two positions in the cavity area as seen in both orange and cyan areas. **b**, $FUCE = 0.24$ and the $L_{3,-}$ cavity mode (same as in Fig. 3c). Only the $B_1$ mode is visible at the edge of the cavity (around $y = -D/2$) as the continuous line is maximum and the dashed line is zero. Slightly after the center of the cavity (orange region), both continuous and dashed lines have the same amplitude. This is confirmed in the near field where both $B_1$ and $E_{2a}$ components are clearly visible. **c**, $FUCE = 0.13$ and the $U_{2.67,-}$ cavity mode (same as in Fig. 4g). At the edge of the cavity (around $y = -D/2$), there is a mixing of the $B_1$ and $E_{2a}$ modes. In the orange region, only the $E_{2a}$ component is visible as the dashed line is maximum, and the continuous line is zero. A similar mixing occurs in the near-field of all higher order modes presented in Fig. 3 and 4 and in open-Dirac potentials in general.